\documentclass[pra,twocolumn,superscriptaddress]{revtex4-2}

\usepackage[english]{babel}
\usepackage{lmodern}
\usepackage{graphicx}
\usepackage{amsmath}
\usepackage{amsfonts}
\usepackage{amssymb}
\usepackage{amsthm}
\usepackage{mathtools}
\usepackage{dsfont}
\usepackage{microtype}
\usepackage{tempora}
\usepackage[usenames]{color}
\usepackage{colortbl}
\usepackage{xcolor}

\usepackage{hyperref}
\hypersetup{colorlinks=true, linkcolor=blue, citecolor=blue, urlcolor=blue}

\begin{document}

 \def\BE{\begin{equation}}
 \def\EE{\end{equation}}
 \def\BA{\begin{array}}
 \def\EA{\end{array}}
 \def\BEA{\begin{eqnarray}}
 \def\EEA{\end{eqnarray}}
 \def\ra{\rangle}
 \def\nn{\nonumber}

 \newcolumntype{C}[1]{>{\centering\let\newline\\\arraybackslash\hspace{0pt}}m{#1}}

\title{Measurement-assisted non-Gaussian gate for Schr{\"o}dinger cat states preparation:
Fock resource state versus cubic phase state}
\author{A.~V.~Baeva}
\affiliation{Saint Petersburg State University, 7/9 Universitetskaya nab., Saint Petersburg, 199034 Russia}
\author{N.~G.~Veselkova}
\affiliation{Peter the Great St.Petersburg Polytechnic University (SPbPU), St.Petersburg, Polytechnicheskaya, 29, 195251, Russia}
\author{N.~I.~Masalaeva}
\affiliation{Institut f\"ur Theoretische Physik, Universit\"at Innsbruck, Technikerstra{\ss}e~21a, A-6020~Innsbruck, Austria}
\author{I.~V.~Sokolov}
 \email{i.sokolov@mail.spbu.ru, sokolov.i.v@gmail.com}
\affiliation{Saint Petersburg State University, 7/9 Universitetskaya nab., Saint Petersburg, 199034 Russia}

\begin{abstract}

In this paper, we consider the preparation of Schr{\"o}dinger cat states using a measurement-assisted gate based on the Fock resource state, the quantum non-demolition (QND) entangling operation, and the homodyne measurement. Previously we have investigated the gate, which for the same goal uses the ancillary non-Gaussian cubic phase state generated from quadrature squeezed states at realistic (finite) squeezing. It is of evident interest to compare the efficiency of both schemes, that is, their ability to produce cat-like superpositions with high fidelity and probability of success. We introduce, in parallel with the exact theoretical description of the gate operation, a clear visual interpretation of the output state based on the semiclassical mapping of the input field variables. The emergence of the superpositions of ``copies'' of the input state in both schemes is due to the fact that such mapping is compatible with two (or, in general, more) sets of values of the output field observables. We demonstrate that even fine details of the output of both gates are effectively predicted and interpreted in our approach. We examine the fidelity and success probability and reveal the ranges of physical parameters where the Fock state-based and the cubic phase state-based gates demonstrate comparable fidelity and (or)  probability of success.

\end{abstract}

\maketitle

\section{Introduction}

Modern quantum technologies require to go beyond the domain of Gaussian states and introduce the non-Gaussian elements~\cite{Walschaers2021}. The non-Gaussian states and non-Gaussian operations~\cite{Bartley2015, Bartley2013, Hu2019, Mardani2020, Liu2022} are crucial for a variety of quantum information processing protocols~\cite{Zhuang2018} related to quantum key distribution~\cite{Guo2019, Ye2019, Kumar2019, Hu2020}, entanglement distillation~\cite{Braunstein2005, Weedbrook2012, Adesso2014}, quantum error correction~\cite{Niset2009, Ofek2016, Vasconcelos2010, Weigand2018, Hastrup2020}, quantum metrology~\cite{Birrittella2012, Carranza2012, Braun2014, Ouyang2016, Zhang2021, Joo2011, Facon2016}, optimal cloning~\cite{Cerf2005}, continuous variable (CV) quantum computing ~\cite{Lloyd1999, Bartlett2002, Mari2012}, and quantum computation on cluster states~\cite{Ohliger2010, Menicucci2006}. In addition, the non-Gaussian states and operations have been shown to be capable of improving the quality of entanglement~\cite{Navarrete-Benlloch2012} and enhancing the accuracy of CV quantum teleportation~\cite{Opatrn'y2000, Yang2009, Xu2015, Wang2015, Hu2017, Kumar2023, Zinatullin2023}. 

The CV Gaussian and non-Gaussian measurement-induced quantum networks are rapidly developing fields of quantum information science and technologies including optical quantum computing~\cite{Takeda2017}, methods for building neural networks on quantum computers~\cite{Killoran2019}, and secure quantum communication~\cite{Lee2019, Guo2017}. 
Optical CV entangled quantum networks~\cite{Walschaers2023} that allow to simulate complex network structures~\cite{Nokkala2018, Sansavini2019}, including quantum communication in the global-scale quantum internet~\cite{Cai2017, Arzani2019}, are an essential resource for measurement-based quantum protocols. To perform quantum protocols, such networks should make use of the non-Gaussian statistics. 

Due to their importance in quantum information science and technology, the non-Gaussian states, such as Fock states~\cite{Ourjoumtsev2006Quantum,Cooper2013, Bouillard2019, Tiedau2019}, Schr{\"o}dinger-cat states~\cite{Ourjoumtsev2006,Ourjoumtsev2007, Quesada2019, Takase2021}, N00N states~\cite{Boto2000, Lee2002}, and the non-Gaussian operations including photon-number subtraction and addition~\cite{Parigi2007, Fiura'sek2009, Marek2008, Kitagawa2006, Namekata2010, Fiura'sek2005, Wakui2007, Wang2015}, the Kerr nonlinearity~\cite{Nemoto2004}, cubic-phase gate~\cite{Gottesman2001, Arzani2017}, sum-frequency generation~\cite{Zhuang2017}, photon-added Gaussian channels~\cite{Sabapathy2017}, and other operations~\cite{Wenger2004} are being intensively studied theoretically and implemented experimentally~\cite{Baragiola2019}.

It is common to call a superposition of macroscopically (classically) distinguishable states of an object~\cite{Haroche2013,Frowis2018} a Schr{\"o}dinger cat state. In quantum optics, a Schr{\"o}dinger cat state is usually related to a superposition of coherent states of an electromagnetic field and is a non-Gaussian state. The cat-like quantum superpositions have been of particular interest since the creation of quantum theory: they serve as a resource for the tests of the foundations of quantum mechanics~\cite{Brune1996,Wenger2003,Garcia2004} and also play a substantial role in up-to-date quantum technologies, including quantum information processing~\cite{Gilchrist2004,Ourjoumtsev2006,Vlastakis2013,Jouguet2013} and quantum computing~\cite{Lloyd1999,Ralph2003,Lund2008,Mirrahimi2014,Cochrane1999}, quantum communication and quantum repeaters~\cite{vanLoock2008, Sangouard2010, Goncharov2022}, quantum teleportation~\cite{Enk2001}, etc., as well as the measurements with non-classical states of light~\cite{Tan2019}.

The generation of large-amplitude coherent-state superpositions ($|\alpha|\geqslant 2$) attracts significant practical interest~\cite{Gilchrist2004, Ralph2003, Walmsley2015} since such cat states can be used as a basis for preparing qubits in CV quantum computation~\cite{Cochrane1999, Ralph2003}, and as a resource for quantum coding with error correction~\cite{Vasconcelos2010, Weigand2018, Hastrup2020}. 

In general, CV Schr{\"o}dinger cat states can be prepared through a strong enough nonlinear interaction using unitary evolution~\cite{Yurke1986}. However, the development of realistic schemes for preparing optical Schr{\"o}dinger cat states with a large number of photons and controlled quantum characteristics is still a challenging task. A variety of well-established approaches were proposed for this goal, including schemes based on photon-number measurement and subtraction~\cite{Dakna1997,Dong2014, Gerrits2010,Asavanant2017,Takahashi2008, Neergaard2006,Neergaard-Nielsen2010, Bashmakova2023}, quantum state engineering~\cite{Bimbard2010, Yukawa2013, Ulanov2016, Huang2015}, iterative growth methods~\cite{Etesse2015,Sychev2017,Lund2004,Laghaout2013}, and other methods for the conditional generation of cat states~\cite{Ourjoumtsev2006Quantum,Ourjoumtsev2007,Gerrits2010,Ourjoumtsev2009,Jeong2006}, some of which have been successfully implemented~\cite{Ourjoumtsev2006,Ourjoumtsev2007,Takahashi2008,Neergaard2006,Etesse2015,Ourjoumtsev2009,Jeong2006}.

By now, the most successful optical method that allows to generate non-classical states of light is the application of intense laser fields to atomic gas~\cite{Lamprou2020, Lewenstein2021, Stammer2022, Rivera-Dean2022}. By using this approach, the cat state containing $\langle n\rangle\thickapprox$ 9.4 photons was generated in the experiment~\cite{Rivera-Dean2022}, which is the largest optical cat state to date.

While most of the schemes proposed and implemented by now rely on such non-Gaussian operations as photon-number measurement and subtraction, the use of the non-Gaussian resource states provides an actual alternative to other cat-states generation methods. An auxiliary resource channel can be prepared in the Fock state~\cite{Ourjoumtsev2007,Etesse2015,Jeong2006,Etesse2014,Sychev2019}, or even in complex superpositions which arise in the schemes based on iterative cat breeding. The generation of optical Schr{\"o}dinger cat states using a photon number state as a resource, a beamsplitter as an entangling element, and homodyne detection was experimentally demonstrated in a low-photon regime~\cite{Ourjoumtsev2007,Etesse2015}.

In previous papers~\cite{Sokolov2020, Masalaeva2022, Baeva2023}, we demonstrated that one can implement a CV measurement-induced two-node logical gate using a quantum non-demolition (QND) entangling operation instead of a beamsplitter, and a cubic phase state as an alternative non-Gaussian resource. The gate conditionally generates a Schr{\"o}dinger cat state: a superposition of two well-spaced on the phase plane ``copies'' of the initial state of the target oscillator (including not only the vacuum state but also other possible initial states). We also proposed a simple visual interpretation of the emergence of cal-like states in similar schemes and their physical properties. Hence, there arises a need to compare different schemes and assess the quantum fidelity and the probability of conditional generation of the output cat-like states.

In this paper, we compare the overall efficiency of the measurement-assisted two-node gate, which is based on the QND entangling operation and the homodyne measurement, for two non-Gaussian resource states of the ancillary oscillator: the Fock resource state versus cubic phase state.

We discuss the similarities and differences between both gates and demonstrate that even fine details of the output state of both schemes are effectively predicted and interpreted not only by an exact theoretical description but also by a visual interpretation in terms of a semiclassical in-out mapping of the quadrature amplitudes of two oscillators. We consider the fidelity and success probability for both schemes, and reveal the ranges of physical parameters where these gates demonstrate comparable efficiencies.
 

\section{Engineering cat-like states using Fock states as a non-Gaussian resource}

\begin{figure}[t!]
\centering
\includegraphics[width=0.6\columnwidth]{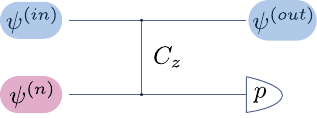}
\caption{Schematic of the protocol for producing cat states using an ancilla oscillator in the Fock state. The states $\psi^{(\text{in})}$ and $\psi^{(n)}$ of the target and ancillary oscillators are sent to the input of the scheme. After applying the $C_z$ entangling operation, the ancilla momentum is measured by a homodyne detector. Depending on the measurement outcome $y_m$, the target oscillator state collapses to a cat-like state $\psi^{(\text{out})}$.}
\label{fig_Scheme_Fock}
\end{figure}

The proposed scheme for producing cat states is depicted in Fig.~\ref{fig_Scheme_Fock}. The input state of the target oscillator can be arbitrary with the wave function in the coordinate representation
\BE
|\psi_1\ra = \int dx_1\psi^{(\text{in})}(x_1)|x_1\ra,
\EE
while the ancillary oscillator is prepared in the Fock state with $n$ photons
\BE
|\psi_2\ra = \int dx_2\psi^{(n)}(x_2)|x_2\ra, 
\EE
where 
\BE
\psi^{(n)}(x_2) = \frac{1}{\pi^{1/4}\sqrt{2^n n!}} H_n(x_2)e^{-x_2^2/2}.
\EE
Here $H_n(x_2)$ is the Hermite polynomial. 

Note that as far as the semiclassical mapping described below does not depend on the input state of the target oscillator and works for a wide range of input states that occupy a limited area on the phase plane, the gate operation is not limited to the vacuum state at the input.

After applying the $C_Z$ entangling operator $\exp(iq_1q_2)$ to the oscillators, the resulting wave function takes the form 
\begin{align}
\label{entangled}
|\psi_{12}\ra &= \frac{1}{\pi^{1/4}\sqrt{2^n n!}}\int dx_1dx_2\psi^{(\text{in})}(x_1)H_n(x_2)e^{ix_1 x_2}\nn\\
& \times e^{-x_2^2/2}|x_1\ra|x_2\ra.
\end{align}

The momentum of ancilla is measured by the homodyne detector with the measurement outcome $y_m$, and the output wave function of the target oscillator is 
\begin{equation}
\label{psi_out}
\psi^{(\text{out})}(x,y_m) = \frac{1}{\sqrt{N}} \psi^{(\text{in})}(x) \big[F\psi^{(n)}\big](y_m - x),
\end{equation}
where $x_1 \to x$ for brevity, $N$ is the normalization factor, and $F$ denotes the Fourier transform,
 \BE
 \label{to_momentum}
[F\psi](y) \equiv \frac{1}{\sqrt{2\pi}}\int dx e^{-iyx} \psi(x).
 \EE
The Fourier transform of the Fock state of ancilla is given by
\begin{align}
\label{Fock_to_momentum}
\left[F\psi^{(n)}\right]\!(y) &=  \frac{1}{\sqrt{2\pi}}\int dx e^{-iyx} \psi^{(n)}(x) = (-i)^n\psi^{(n)}(y) \nn\\
& = \frac{(-i)^n }{\pi^{1/4}\sqrt{2^n n!}} H_n(y)e^{-y^2/2},
\end{align}
and we arrive at the output wave function of our gate, 
 \BE
 \label{psi_out_norm}
\psi^{(\text{out})}(x,y_m) = \frac{1}{\sqrt{N}}\tilde\psi^{(\text{out})}(x,y_m). 
 \EE
Here $\tilde\psi^{(\text{out})}(x,y_m)$ is the non-normalized output wave function collapsed by the measurement,
\begin{align}
 \label{psi_out_red}
\tilde\psi^{(\text{out})}(x,y_m) & = \psi^{(\text{in})}(x)\nn\\
&\times \frac{(-i)^n}{\pi^{1/4}\sqrt{2^nn!}}H_n(y_m - x)e^{-(y_m - x)^2/2}.
\end{align}

As it will be demonstrated later, the considered above scheme produces at the output a superposition of two copies of the initial state of the target oscillator, by analogy with the gate using an ancilla oscillator in the cubic phase state~\cite{Sokolov2020,Masalaeva2022,Baeva2023}. Since the superpositions of coherent states are widely considered as a tool for the error correction schemes, and as a logical basis for computational operations~\cite{Cochrane1999, Ralph2003}, in the following we will focus on the vacuum initial state of the target oscillator.

\section{Schr{\"o}dinger cat state generation in semiclassical picture}
\label{sec_semicl}

As we have already discussed in Ref.~\cite{Sokolov2020,Masalaeva2022,Baeva2023}, for the gate using the cubic phase state as the non-Gaussian resource, the closest to the exact solution representation of the output state in the form of a cat-like superposition of two ``copies'' of the target state can be effectively evaluated in the semiclassical picture. For the Fock state-based gate, the semiclassical in-out mapping of the quadrature amplitudes of the target state is also easily accessed starting from the Heisenberg picture. 

\begin{figure}[t!]
\centering
\includegraphics[width=0.95\columnwidth]{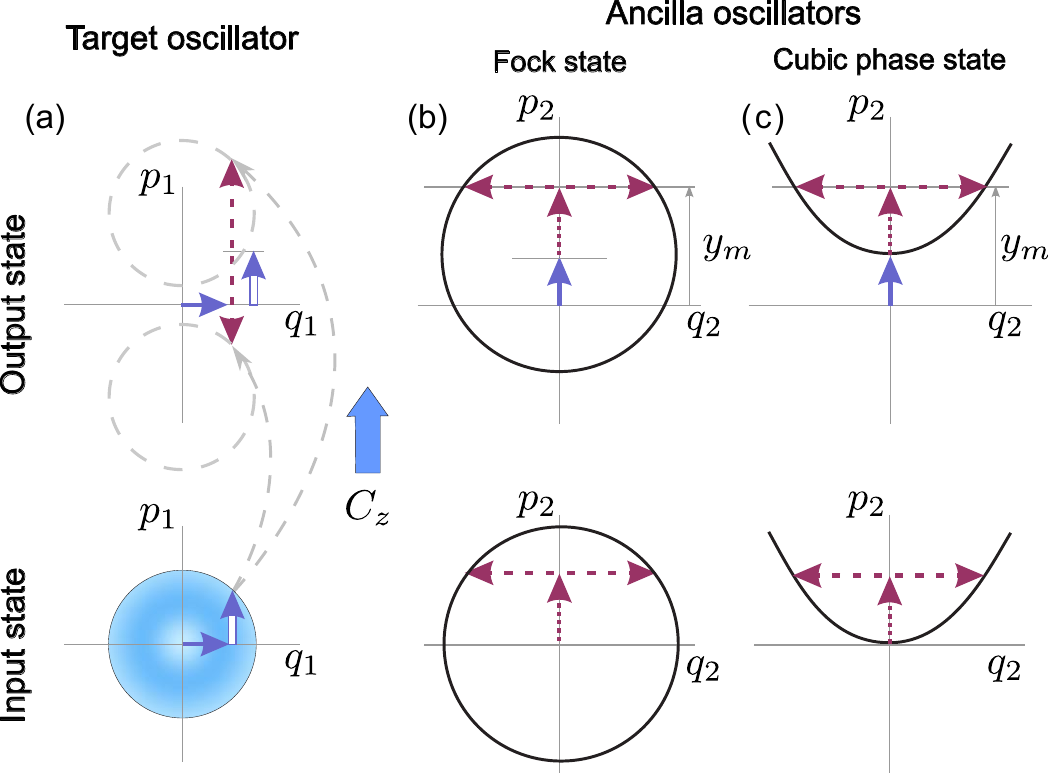}
\caption{The measurement-induced semiclassical mapping of the quadrature amplitudes of the target [(a) column] and ancilla oscillators [(b) and (c) columns]. Here, column (b) represents the gate with the ancillary oscillator in the Fock state $\psi^{(n)}$, while column (c) corresponds to the gate with the ancillary oscillator prepared initially in a perfect cubic phase state. The randomly chosen initial amplitudes undergo the entangling non-demolition $C_z$ operation, which transfers the target oscillator amplitudes to the ancilla and vice versa. The measurement of ancilla momentum with the outcome $y_m$ [top row of (b) and (c) columns] specifies the values of the amplitudes compatible both with the preparation and the measurement. Since these amplitudes in both cases arise as multivalued, the target oscillator state collapses to a cat-like state [the top row of (a) column]. The parameters of both gates can be matched in such a way that the arising cat states may have similar spacing between the copies but different probabilities of success and fidelities of the state preparation.}
\label{fig_Scheme}
\end{figure}

The semiclassical approach that we develop here can be described as follows. We consider the transformation of the canonical variables of both fields performed by the entangling operation in the Heisenberg picture. The input Fock resource state is specified by its energy, which in the semiclassical approximation imposes the relation on its canonical variables treated as c-numbers. The measurement of the ancilla canonical variable (momentum) is also described semiclassically by simply substituting momentum with its observed value. This results in an explicit relation between the target oscillator input and output canonical variables, which represents the in-out mapping performed by the gate in the semiclassical approximation. The obtained relation is used for the reconstruction of the target oscillator wave function. This consideration in a simple and visual form demonstrates that two-headed (or even more complex) Schr{\"o}dinger cat arises when the measurement outcome is compatible with not one but with two (or more) values of the target oscillator variables.

We define the coordinate $q_i$ and momentum operators $p_i$ as $a_i=(q_i+ip_i)/\sqrt{2}$, where $[q_i,p_i]=i$. As far as the initial ancilla state is the Fock state $|n\rangle$, we assume that the basic equation that specifies the Fock state 
\BE
\frac{1}{2}(q^2_2 + p^2_2)|n\rangle = (n + \frac{1}{2})|n\rangle, 
\EE
in the semiclassical approximation yields 
 $$
 \big(q_2^{(\text{in})}\big)^2 + \big(p_2^{(\text{in})}\big)^2 = 2n + 1.
 $$
The two-mode entangling QND operation $C_z\sim\exp(iq_1q_2)$ applied to the initial state of the oscillators performs the following transformation in the Heisenberg picture, 
\begin{gather}
   q'_1 = q_1^{(\text{in})},\quad p'_1 =p_1^{(\text{in})}+q^{(\text{in})}_2,\nn\\
    q'_2 = q^{(\text{in})}_2,\quad p'_2 = p^{(\text{in})}_2 + q_1^{(\text{in})}.\nn
\end{gather}
The measurement of the ancillary oscillator momentum with the outcome $y_m$ yields $p^{\prime}_2 \to y_m$ in the semiclassical approximation, and we arrive at $q'_2 \to \pm\sqrt{2n+1-(y_m-q^{(\text{in})})^2}$. The in-out semiclassical mapping of the target oscillator quadrature amplitudes takes the form
 \BE
 \label{Fock_cat_transf}
 \BA{ll}q_1^{(\text{out})} = q_1^{(\text{in})},\\
 p_1^{(\text{out})} = p_1^{(\text{in})} \pm \sqrt{2n + 1 - (y_m - q_1^{(\text{in})})^2}.
 \EA
 \EE
As one can see, two solutions for the momentum arise. The latter indicates that the output state of the gate is a superposition of two macroscopically distinguishable contributions - a cat-like state. The measurement adds to their momenta the quantities
\begin{equation}
   \label{mom_displ}
\pm \delta p(x) = \pm \sqrt{2n + 1 - (y_m - x)^2},  
\end{equation}
where $x$ stands for $q_1^{(\text{in})}$ for the shorthand. In order to find the corresponding components of the output wave function we conjecture, in analogy to Ref.~\cite{Sokolov2020,Masalaeva2022,Baeva2023}, that for a given component of the target oscillator wave function the semiclassical factor added by the measurement can be defined as 
 \BE
 \label{semicl}
\varphi_{\text{scl}}^{(\pm)}(x) \sim \sqrt{P^{(\pm)}(x)}\exp\!\left[\pm i\!\int dx \delta p(x)\right],
 \EE
up to the phase factor that will be discussed later. Here $P^{(\pm)}(x)$ is the partial probability density of the measurement outcome $y_m$ for a given $x$, and the exponential factor provides the displacement in momentum implied by Eq.~\eqref{Fock_cat_transf}.

Consider the semiclassical density distribution of the ancillary oscillator in the phase plane shifted by $x$ along the momentum quadrature by the entangling $C_z$ operation, as illustrated in Fig.~\ref{fig_Scheme}(b),
 $$
P^{\text{(anc)}}(q_2,p_2,x) = \frac{1}{\pi}\delta\left[(p_2 - x)^2 + q_2^2 - (2n + 1)\right], 
$$
where
$$
\int dq_2\,dp_2 P^{\text{(anc)}}(q_2,p_2,x) = 1.
 $$

 \begin{figure*}[t!]
\centering
\includegraphics [width=0.9\textwidth]{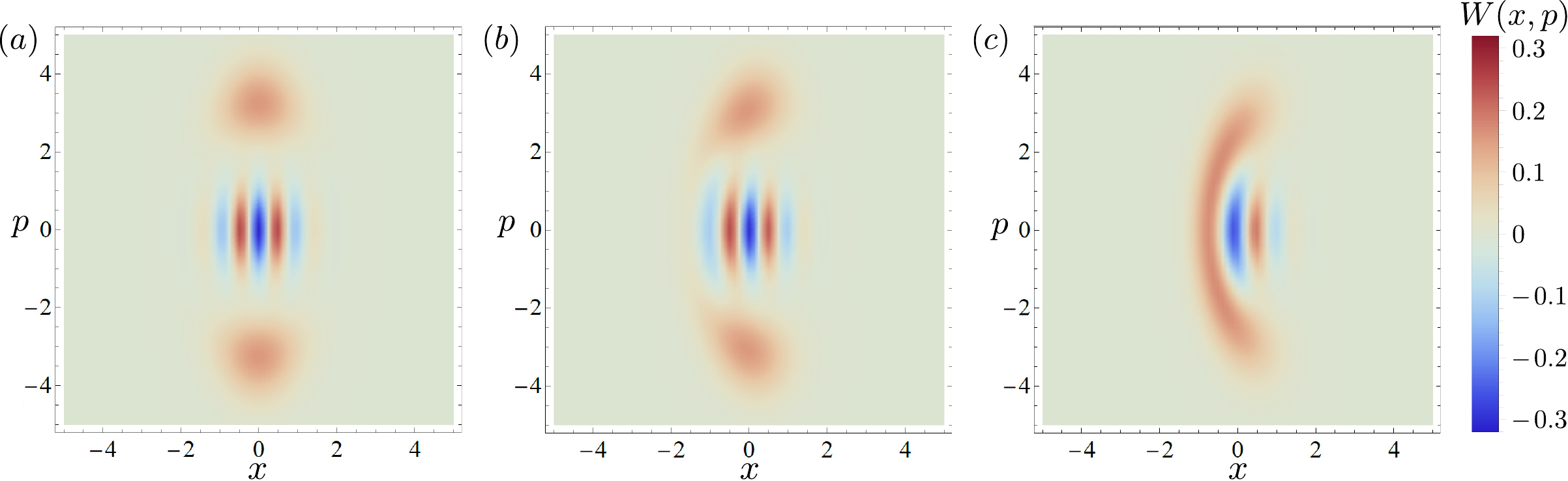}
\caption{The Wigner function of the output state for the vacuum input state and the Fock resource state with $n=5$ for different outcomes of the momentum measurements of ancilla: (a) $y_m = 0$, (b) $y_m = 1$, and (c) $y_m = 2$.}
\label{fig_Wigner_Fock}
\end{figure*}

The probabilities $P^{(\pm)}(x)$ are defined through the overlap of $P^{\text{(anc)}}(q_2,p_2,x)$ with the area on phase plane that corresponds to the measurement outcome $y_m$, 
\begin{gather}
 P^{(+)}(x) + P^{(-)}(x) \sim \int dq_2\,dp_2\delta(p_2 - y_m)P^{\text{(anc)}}(q_2,p_2,x)\nn\\
 =  \frac{1}{\pi}\int dq_2\,dp_2\delta(p_2 - y_m)\delta\left[(p_2 - x)^2 + q_2^2 - (2n + 1)\right]\nn\\
 = \frac{1}{\pi}\int dq_2 \delta\Big\{\big[q_2 - \Delta(x)\big]\big[q_2 + \Delta(x)\big]\Big\} = \frac{1}{\pi\Delta(x)}.\nn
\end{gather}
Here $\Delta(x) = \sqrt{2n + 1 - (y_m - x)^2} > 0$ in the regime where the mapping shown in Fig.~\ref{fig_Scheme}(b) is multivalued. If the coordinates of the target state are localized within some area, $|x| \leq \delta x$, one can expect a ``good'' cat state for such measurement outcomes, that $y_m\pm\delta x <\sqrt{2n+1}$, since for all $x$ there are two points of overlap, as depicted in Fig.~\ref{fig_Scheme}(b).
Note that $\Delta(x) = \delta p(x)$, hence,
\BE
\label{weights}
P^{(\pm)}(x) \sim \frac{1}{|\delta p(x)|}.
\EE

By inserting~\eqref{weights} into~\eqref{semicl}, we arrive at a semiclassical added factor $\varphi_{\text{scl}}(x)$ that looks similar to the semiclassical wave function of a stationary state expressed through the momentum $\delta p(x)$. The superposition of semiclassical added factors, see Eq.~\eqref{semicl}, can be evaluated and written as
 \BE
 \label{varphi_scl}
\varphi_{\text{scl}}(n,z) \sim \frac{1}{\big(1 -  z^2\big)^{1/4}}
\big[e^{i\phi(n,z)} + (-1)^n e^{-i\phi(n,z)}\big],
\EE
where
\begin{gather}
\label{z_phi}
z = \frac{x - y_m}{\sqrt{2n + 1}}, \\
\phi(n,z) = \frac{1}{2}(2n + 1)\left(z\sqrt{1 - z^2} +
\arcsin z\right).\nn
\end{gather}
The relative phase between two contributions $(-1)^n$ is chosen in accordance with the exact solution, see Eq.~\eqref{psi_out_red}: for even $n$ the added factor reaches a maximum at $x - y_m = 0$, and for odd $n$ it is equal to 0.

To be specific, consider the action of the gate on the vacuum input state. An approximate version of Eq.~\eqref{varphi_scl} arises when the wave function of the input state is concentrated in a relatively small area of coordinate near $x=0$. Assuming the denominator in~\eqref{varphi_scl} to be constant and Taylor expanding the phase $\phi(n,z)$ up to a linear term in the coordinate $x$, we receive
 \BE
 \label{phase_approx}
\phi(n, z) \approx  \theta + p^{(+)}x + \ldots,
 \EE
where
 \BE
 \label{momentum_shift}
\theta = \phi\left(n, -y_m/\sqrt{2n + 1}\right), \quad p^{(+)} = \sqrt{2n + 1 - y_m^2}.
 \EE
In the limit $y_m \to 0$ one has $\theta \to 0$. As seen from Eq.~(\ref{varphi_scl}), when the measured momentum of the ancilla oscillator is small, one can expect an even/odd cat-like output superposition depending on the number of photons in the resource state. For the vacuum input state $\psi^{(0)}(x)$, the displaced component $\exp(ip^{(+)}x)\psi^{(0)}(x)$ of the output state is the Glauber coherent state $|\alpha\ra$, where $\alpha=ip^{(+)}/\sqrt{2}$.

The approximations made up to now (that is, the semiclassical picture and linearization) yield the gate output state in the form of a superposition of two symmetrically displaced along the momentum axis coherent states,
 \BE
 \label{out_vac_semicl}
|\psi^{\text{(coh)}}\ra =  \frac{1}{\sqrt{{\cal N}_{\text{vac}}}}(e^{i\theta}|\alpha\ra + (-1)^n e^{-i\theta}|-\alpha\ra),
 \EE
where the normalization factor is
 $$
{\cal N}_{\text{vac}} = 2\big[1 + (-1)^n\cos(2\theta) e^{-2|\alpha|^2}\big].
 $$
Note that the relative phase $\theta$ for an arbitrary measurement outcome may acquire values that do not correspond to an even or odd cat state.

In the coordinate representation, the state~\eqref{out_vac_semicl} looks like
 \BE
 \label{cat_Glauber_coord_even}
\psi^{\text{(coh)}}(x,y_m) = \frac{\sqrt{2}}{\pi^{1/4}}
\frac{\cos\big(\theta + p^{(+)} x\big)}{\sqrt{1 + \cos(2\theta) \exp(-{p^{(+)}}^2)}} e^{-x^2/2},
 \EE
for the even $n$'s, and
 \BE
 \label{cat_Glauber_coord_odd}
\psi^{\text{(coh)}}(x,y_m) = i\frac{\sqrt{2}}{\pi^{1/4}}
\frac{\sin\big(\theta + p^{(+)} x\big)}{\sqrt{1 - \cos(2\theta) \exp(-{p^{(+)}}^2)}} e^{-x^2/2},
 \EE
for the odd ones respectively. As can be seen from Eq.~\eqref{out_vac_semicl}, for the optimal measurement outcome $y_m = 0$, when $\theta = 0$, the even or odd cat state can be produced as far as the approximations made in the semiclassical picture are valid. 

In the following, we will evaluate fidelity between the exact form of the output state, see Eq.~\eqref{psi_out_red}, and the even/odd cat state derived above. The latter will help us to match the output state with a ``perfect'' even/odd superposition of coherent states. 


\section{Wigner function of cat states}

The Wigner function of the exact output state~\eqref{psi_out_red} is given by
\begin{gather}   
W^{(\text{out})}(x,y) =
\frac{1}{\pi}\int dz\,\psi^{(\text{out})*}(x+z,y_m)\psi^{(\text{out})}(x-z,y_m)\,\nn\\
\times e^{2iyz}= \frac{1}{\pi N}\int dz \psi^{(\text{in})*}(x+z)\psi^{(\text{in})}(x-z)\nn\\
\times \big[F\psi^{(n)}\big]^*(y_m-(x+z)) \big[F\psi^{(n)}\big](y_m-(x-z)) e^{2iyz}.
\label{Wigner_out}
\end{gather}
In Fig.~\ref{fig_Wigner_Fock} we present a set of the Wigner functions of the output state in dependence on the measured ancilla momentum $y_m$ for the vacuum input state of the target oscillator, $\psi^{(\text{in})}(x) = \psi^{(0)}(x)$. The ancilla oscillator is chosen to be in the Fock state with $n = 5$, which for the optimal measurement outcome, $y_m = 0$, provides the displacements of the copies equal to $\pm \sqrt{2n + 1} \approx \pm 3.32$.

For the measurement outcome $y_m = 0$, the measurement-induced momentum displacements in the output state, see Eq.~\eqref{mom_displ}, are almost independent of the position $x$ of the initial point, chosen within the support region of the target oscillator in the phase plane. In other words, for $y_m = 0$, the mapping is almost insensible to small shifts along the momentum axis of the circle that represents the resource Fock state for a large enough number of photons, due to its geometrical shape, see Fig.~\ref{fig_Scheme_Fock}(b). As a result, one can expect a superposition of two ``good'' copies of the target state at the output of the scheme. As can be seen from Fig.~\ref{fig_Wigner_Fock}(a), the Wigner function of the output state given by Eq.~\eqref{Wigner_out} for $y_m = 0$ is in good agreement with the above discussion.

For large enough measurement outcomes $y_m$, the copies of the input state of the target oscillator suffer from a shearing deformation of the opposite sign, see Fig.~\ref{fig_Wigner_Fock}(b), due to almost linear dependence of displacements on $x$.

\begin{figure}[h!]
\centering
\includegraphics[width=0.9\linewidth]{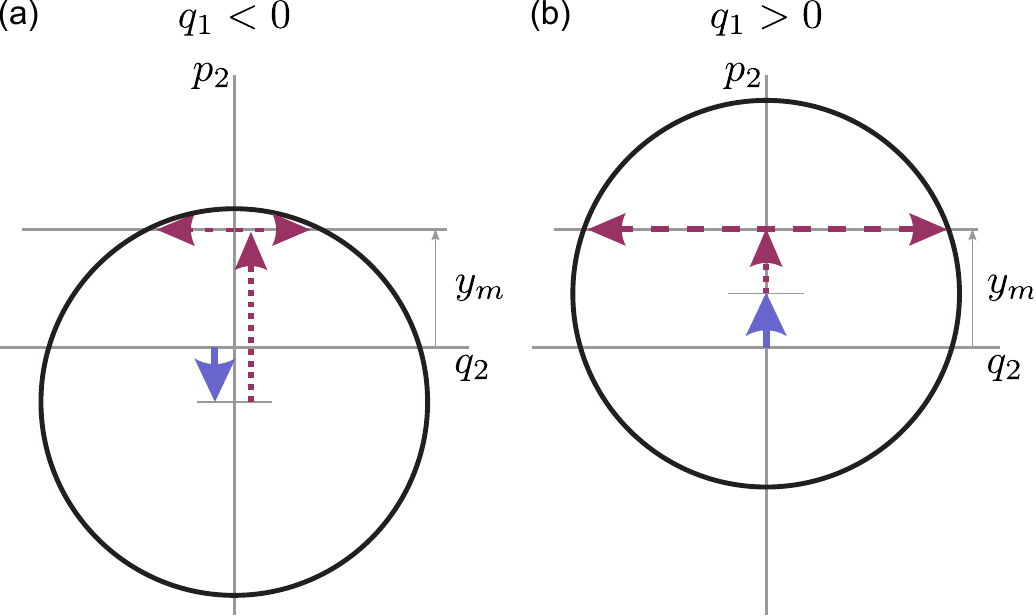}
\caption{The measurement-induced splitting of the ancilla coordinate (dashed horizontal arrows) for different values of the target oscillator coordinate $q_1$ (solid blue arrows) and the same measurement outcome of the ancilla momentum $y_m = 2$.}
\label{fig_Deformation}
\end{figure}

In Fig.~\ref{fig_Deformation}, we illustrate how such shearing deformation arises in our scheme. We take the same parameters, as in Fig.~\ref{fig_Wigner_Fock}(c), that is $y_m = 2$ and the Fock resource state with $n = 5$, and plot the appearing measurement-induced splitting of the ancilla coordinate. The latter determines the displacement of the copies along the momentum axis in the output state, see Eq.~\eqref{mom_displ}. As can be seen from Fig.~\ref{fig_Deformation}, for a given measurement outcome of the ancilla momentum $y_m$, the arising splitting of the ancilla coordinate $q_2$ may essentially depend on the target oscillator coordinate $q_1$. In case of negative values of $q_1$, see Fig.~\ref{fig_Deformation}(a) the displacement along the momentum axis $\delta p(x)$ can turn out to be small, which leads to the merging of the copies, see Fig.~\ref{fig_Wigner_Fock}(c). For even larger negative values of $q_1$, the expression standing under the root in $\delta p(x) = \sqrt{2n + 1 - (y_m - x)^2}$ turns out to be negative and, as a consequence, two crossings of the circle, see Fig.~\ref{fig_Deformation}(a), no longer exist for some $x$. While for the positive $q_1$, see Fig.~\ref{fig_Deformation}(b), the displacement of the copies is more pronounced.

It is also clear from the developed above qualitative picture that one can generate even/odd superpositions with high fidelities out of an arbitrary target state using the Fock resource states with the large enough dimensions of the uncertainty region (that is, with a large enough number of quanta) in comparison with the target state, which makes this cat state generating gate universal.


\begin{figure}[t!]
\centering
\includegraphics[width=0.9\linewidth]{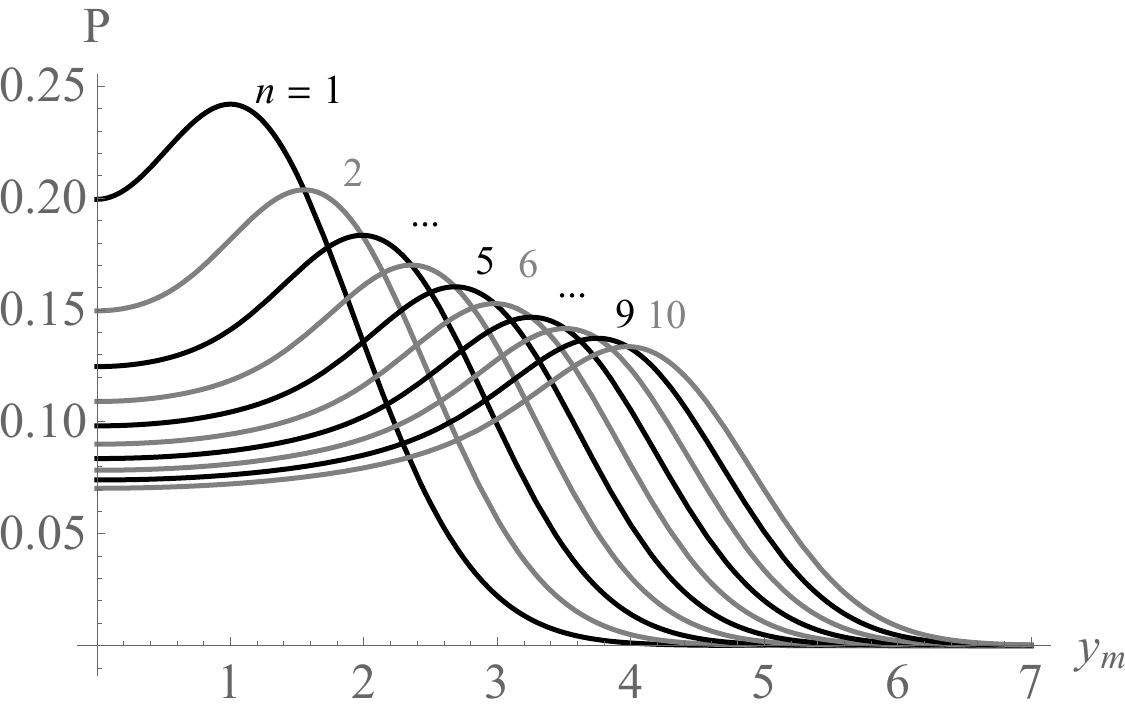}
\caption{The probability density $P$ as a function of the ancilla momentum measurement outcome $y_m$ for the Fock resource states with $n =$ 1,\,2,\ldots 10 photons.}
\label{fig_Probability}
\end{figure}
 
\section{Probability versus fidelity}

The probability density to observe the measurement outcome $y_m$ is given by the norm of the collapsed wave function~\eqref{psi_out_red},
 \BE
 \label{probability_density}
P(y_m)=\langle\tilde{\psi}^{(\text{out})}|\tilde{\psi}^{(\text{out})}\rangle=\int dx |\tilde{\psi}^{(\text{out})}(x)|^2.
\EE

For the vacuum input state of the target oscillator and the Fock resource state with $n$ photons, the probability density can be rewritten as
 \BE
 \label{psi_out_red_unnormalized}
P(y_m)=\frac{1}{\pi2^nn!}\int dx|H_n(y_m - x)|^2e^{-(y_m - x)^2}e^{-x^2}, 
 \EE
and is shown in Fig.~\ref{fig_Probability} for different measurement outcomes.

\begin{figure}[t!]
\centering
\includegraphics[width=0.98\columnwidth]{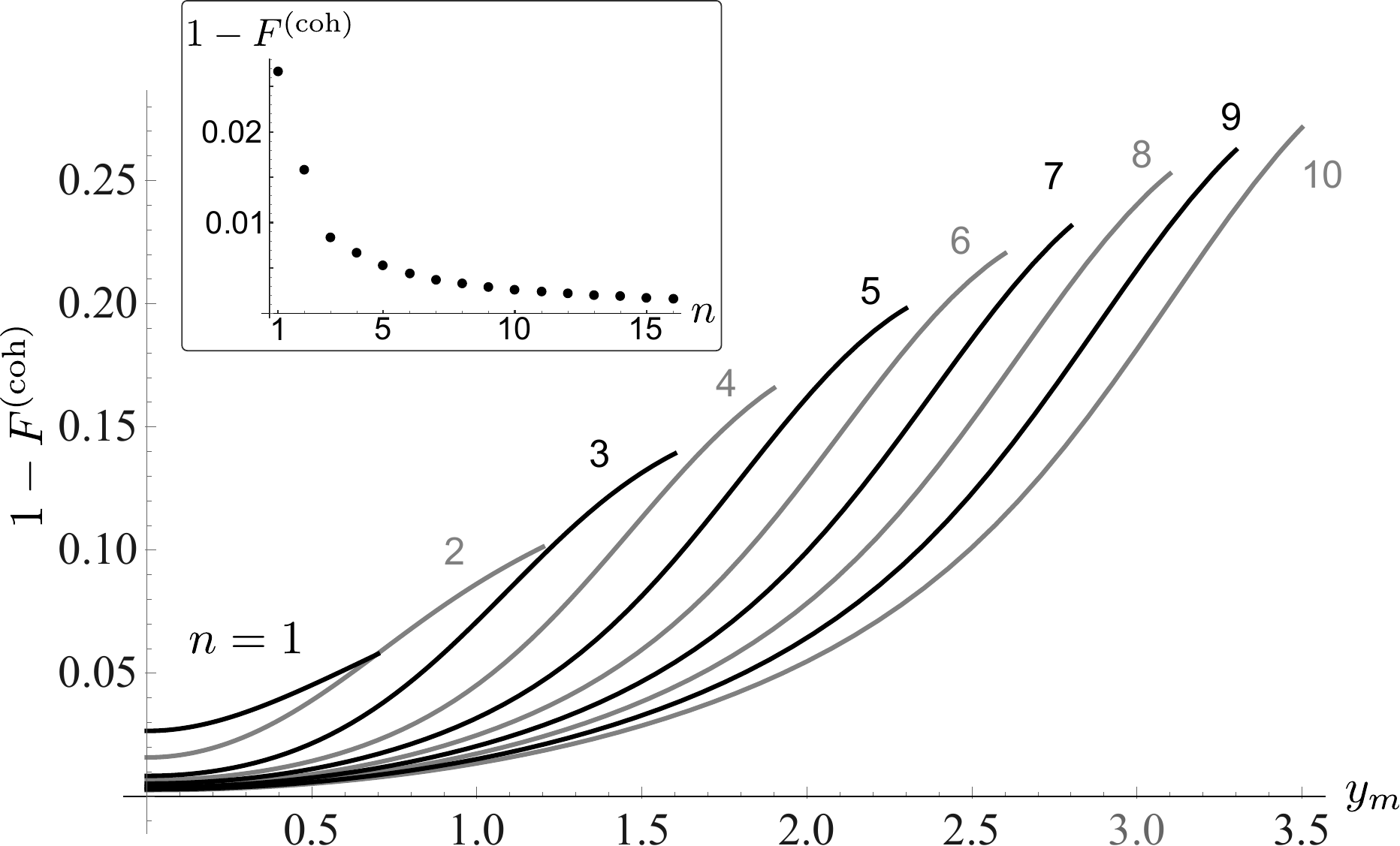}
\caption{Infidelity $1 - F^{\text{(coh)}}$ between the exact solution for the output state [Eq.~\eqref{psi_out_red}] and the corresponding superposition of two undistorted coherent states [Eq.~\eqref{out_vac_semicl}] as a function of the measurement outcome $y_m$. Here, the target oscillator is initially prepared in the vacuum state, and ancilla oscillator is in the Fock state with $n=$1, 2,\ldots 10 photons. For every $n$, the infidelity is plotted within the range of positive $y_m$ satisfying the condition $\Delta(x) > 0$, where one can expect a fair output cat state as it follows from the  visual interpretation of the gate performance, see Fig.~\ref{fig_Scheme}. We plot the infidelity for $y_m\geq 0$ only, since it is an even function of $y_m$. The inset shows the infidelity $1 - F^{\text{(coh)}}$ as a function of the number of photons in the resource state $n$ for the measurement outcome $y_m =0$. As can be seen from the plot, such measurement outcome provides the highest fidelity.}
\label{fig_Infidelity}
\end{figure}

In order to evaluate how close is the output state [Eq.~\eqref{psi_out_norm}] to a superposition of two coherent states, we consider two measures. The first one is the fidelity between the exact output state and the cat-like superposition~\eqref{out_vac_semicl}, where the relative phase $\theta$, given by Eq.~\eqref{momentum_shift}, provides the best match to the exact solution but in general does not correspond to an even/odd cat state,
 \BE
 \label{Fidelity_coh}
F^{\text{(coh)}} = \left|\int dx\,\psi^{(out)*}(x,y_m)\psi^{\text{(coh)}}(x,y_m)\right|^2.
 \EE
The corresponding dependence of the infidelity $1 - F^{\text{(coh)}}$ on the measurement outcome is shown in Fig.~\ref{fig_Infidelity}. For the optimal measurement result $y_m = 0$ the infidelity is depicted in more detail in the inset of Fig.~\ref{fig_Infidelity}.

As seen from Fig.~\ref{fig_Infidelity}, for any $n$ the best fidelity is achieved for $y_m = 0$ where the distortion of copies is minimal, as it follows from the semiclassical picture. The Wigner functions presented in Fig.~\ref{fig_Wigner_Fock} for $n = 5$ also agree with this conclusion.

The dependence of the fidelity on $n$, shown in the inset of Fig.~\ref{fig_Infidelity}, demonstrates that with the increase of the photon number of ancilla, the gate, based on the Fock resource state, can produce superpositions of undistorted coherent states with high fidelities.

\begin{figure}[t!]
\centering
\includegraphics[width=0.8\columnwidth]{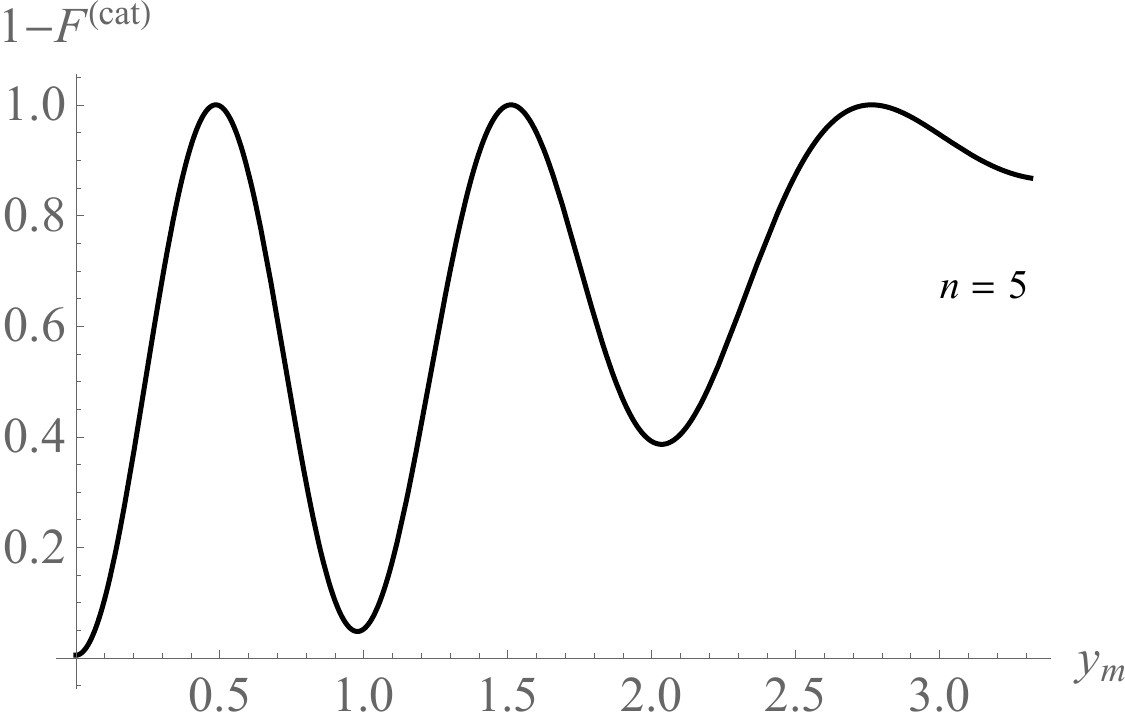}
\caption{Infidelity $1 - F^{\text{(cat)}}$ between the exact solution for the output cat state~\eqref{psi_out_red} and the corresponding odd cat state for the vacuum input state and the resource state with photon number $n=5$~\eqref{cat_Glauber_coord_odd} as a function of the measurement outcome $y_m$.} 
\label{fig_Infidelity_n5}
\end{figure}

Since some error correction protocols rely on the even/odd superpositions of coherent states~\cite{Hastrup2020, Schlegel2022}, we plot in Fig.~\ref{fig_Infidelity_n5} the infidelity $1 - F^{\text{(cat)}}$ between the odd cat state, which is given by Eq.~\eqref{cat_Glauber_coord_odd}, for $n = 5$ and optimal measurement outcome (that is, for $y_m=0$ and $\theta = 0$), and the output state~\eqref{psi_out_red}, when the observed ancilla momentum is not optimal,
 \BE
 \label{Fidelity_odd}
F^{\text{(cat)}} = \left|\int dx\,\psi^{(out)*}(x,y_m)\psi^{\text{(coh)}}(x,0)\right|^2.
 \EE

As the difference between measured momentum $y_m$ and the optimal one increases, the fidelity $F^{\text{(cat)}}$ (for both even and odd cats) degrades rapidly due to the evolution of the relative phase between the copies. The relative phase $\theta$ depends on the measurement result $y_m$, see Eqs.~\eqref{z_phi} and~\eqref{momentum_shift}, and, therefore, can take values from $0$ to $2\pi$, bringing us closer or further away from the desired state, resulting in the oscillating behavior of the infidelity in Fig.~\ref{fig_Infidelity_n5}. Hence, this is mostly a phase effect.

\begin{figure}[t!]
\centering
\includegraphics[width=0.8\columnwidth]{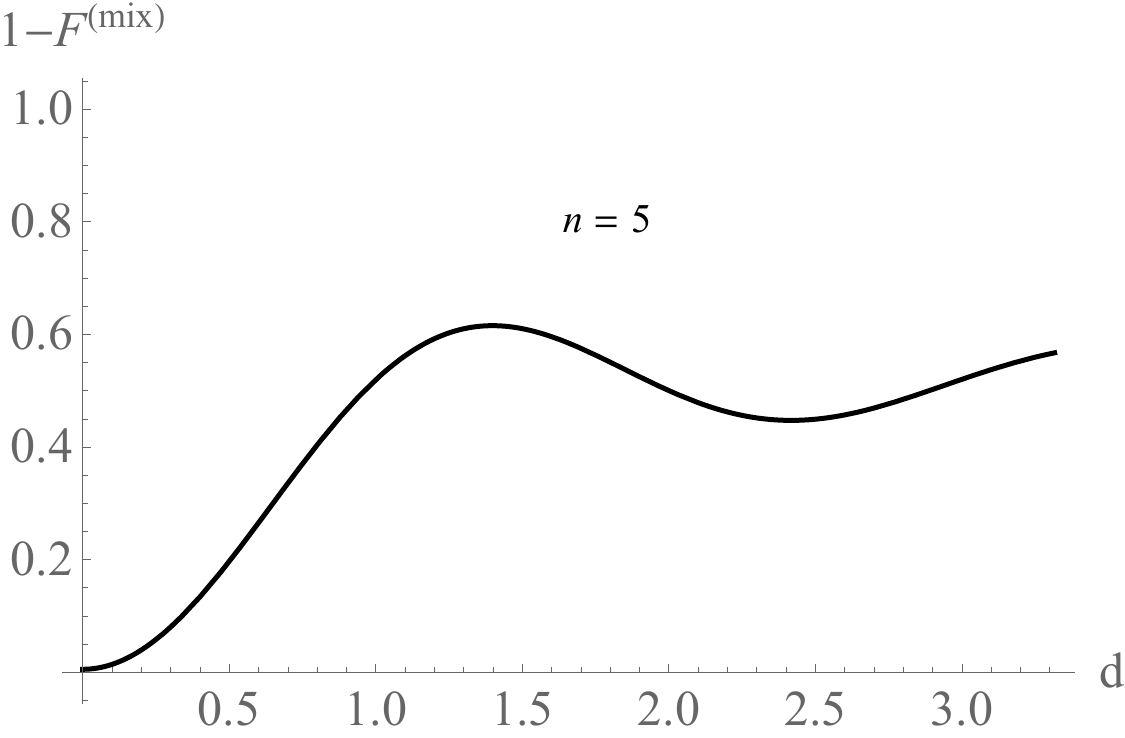}
\caption{Infidelity $1 - F^{\text{(mix)}}$, see Eq.~\eqref{Fidelity_mix}, between the output mixed state and the target odd cat state as a function of the acceptance interval $d$.} 
\label{fig_Infidelity_mix}
\end{figure}

In order to increase the probability of the cat state preparation, one can introduce an acceptance interval $-d/2,\,d/2$ of width $d$ for the measured ancilla momentum, centered at $y_m = 0$. In this case, a mixed state arises as the gate output. The weighted fidelity between the even/odd cat state and the mixed output state is evaluated as
 \BE
 \label{Fidelity_mix}
F^{\text{(mix)}}(d) =
\frac{1}{P^{\text{(mix)}}(d)}\int_{-d/2}^{+d/2} dy_m\,P(y_m)F^{\text{(cat)}}(y_m).
 \EE%
Here
 $$
P^{\text{(mix)}}(d)=\int_{-d/2}^{+d/2} dy_m\,P(y_m),
 $$
is the probability that the measurement outcome will fit the acceptance interval. For the resource state with $n = 5$ photons, the infidelity $1 - F^{\text{(mix)}}$ is shown in Fig.~\ref{fig_Infidelity_mix}. For a small enough acceptance interval, $d \ll \sqrt{2n + 1}$, the probability $P^{\text{(mix)}}(d)$ is evaluated as $P^{\text{(mix)}} \sim P(y_m\to 0)\times d$. This follows from the weak dependence of the probability density on $y_m$ within the interval where the fidelity is high.

\section{Fock resource state versus cubic phase state}

In this section, we compare the efficiency of the gate described above with the previously investigated scheme~\cite{Sokolov2020,Masalaeva2022,Baeva2023}, which uses the cubic phase state as the non-Gaussian resource instead of the Fock state. It is instructive to discuss both gates by illustrating the similarities and differences between both schemes. Their visual representation is shown in Fig.~\ref{fig_Scheme}, where the columns (b) and (c) represent the semiclassical mapping using the Fock resource state and cubic phase state, respectively. If the latter one is prepared from a perfectly squeezed state through quantum evolution with cubic Hamiltonian, its semiclassical support region can be depicted as a parabola.

Since both gates use the same QND operation to entangle the target and the ancillary oscillator, the mutual exchange of quadrature amplitudes looks similar in both devices. The gate based on the cubic phase state is described by a multivalued semiclassical mapping of the form~\cite{Masalaeva2022}
 \BE
 \label{cubic_cat_transf}
 \BA{ll}q^{(\text{out})}_1 = q_1^{(\text{in})},\\
 p^{(\text{out})}_1 = p_1^{(\text{in})} \pm (3\gamma)^{-1/2}\sqrt{y_m - q_1^{(\text{in})} - p_2^{(\text{in})}},\EA
 \EE
which can be derived following the same steps as in the case of the Fock resource state. Here $\gamma$ is the coupling parameter of the evolution operator $\exp(-i\gamma q_2^3)$ used for the preparation of cubic phase state, and $p_2^{(\text{in})}$ is the ancilla momentum before the application of the cubic deformation operator.

Similar to Eqs.~\eqref{Fock_cat_transf}, two solutions arise, which indicates that the output state is a superposition of two macroscopically distinguishable contributions - a cat-like state. The difference in the geometrical shapes of the semiclassical support regions for two resource states, evident from Fig.~\ref{fig_Scheme}, manifests itself in different sensitivity of the output quadrature amplitudes, see Eqs.~\eqref{Fock_cat_transf} and~\eqref{cubic_cat_transf}, to the choice of position of the initial point $\{q_1^{(\text{in})},\,p_1^{(\text{in})}\}$ within the target oscillator support region, and, hence, in different distortion of the copies.

As we have already discussed, the Fock state-based gate prepares a cat-like state such that the spacing between the copies is determined by the number of quanta in the resource state and the momentum measurement outcome $y_m$, where minimal distortion takes place for $y_m = 0$. In order to put in correspondence the output state of the gate based on the cubic phase state, one has to estimate the proper values of the following parameters: (i) the parameter $\gamma$ of cubic nonlinearity; (ii) the measurement outcome of the momentum of the ancillary oscillator $y_m$; and (iii) the factor $s \leq 1$ of initial squeezing of the ancilla momentum before the cubic deformation is applied.

For the vacuum input state of the target oscillator and perfect initial squeezing, $s \to 0$, the spacing between the centers of copies is estimated from Eq.~\eqref{cubic_cat_transf} by setting $q_1^{(\text{in})} \to 0$ and $p_2^{(\text{in})} \to 0$, which gives for the displacement of the copies $\pm p^{(+)} = \pm\sqrt{y_m/(3\gamma)}$. To achieve the same spacing between the copies of the input state of the target oscillator as for the discussed above Fock state-based gate with $n = 5$, where $\pm p^{(+)} = \pm \sqrt{2n + 1} = \pm \sqrt{11}$,  we accept such measurement outcomes, that $y_m = 33\gamma$. 

It can be shown, see Ref.~\cite{Masalaeva2022,Baeva2023}, that the even/odd cat states represented by~\eqref{out_vac_semicl}, where the phase $\theta$ is set to 0, arise for a discrete set of values of the pairs $\{y_m,\,\gamma\}$. Table~\ref{tab_ym_gamma} shows the values that correspond to the odd cat state similar to one produced by the Fock state-based gate for $n = 5$, $y_m = 0$.
\begin{table}[h]
    \centering
	\begin{tabular}{|C{2cm}|C{2cm}|}
    \hline
    $y_{m}$& $\gamma$ \\ \hline
    1.066 & 0.032\\ \hline
    2.486 & 0.075 \\ \hline
    3.907 & 0.118\\ \hline
    5.328 & 0.161\\ \hline
    6.749 & 0.205\\ \hline
    8.170 & 0.248\\ \hline
    9.591 & 0.291\\ \hline
    11.012 & 0.334 \\ \hline
    12.432 & 0.377 \\ \hline
	\end{tabular}
\caption{Values of the parameter $\gamma$ of the cubic nonlinearity and the ancilla momentum measurement outcome $y_m$ that correspond to the odd output cat state with the same displacement of copies along the momentum axis as for the Fock state-based gate considered above.}
\label{tab_ym_gamma}
\end{table}

The Fock state-based gate at $n = 5$ and $y_m = 0$ produces the odd cat state with the probability density $P =  0.098$ and the fidelity $F^{\text{(cat)}}$ such that $1 - F^{\text{(cat)}} = 0.005$ (see Figs.~\ref{fig_Probability} and~\ref{fig_Infidelity}, respectively).

\begin{figure}[t!]
\centering
\includegraphics[width=0.7\columnwidth]{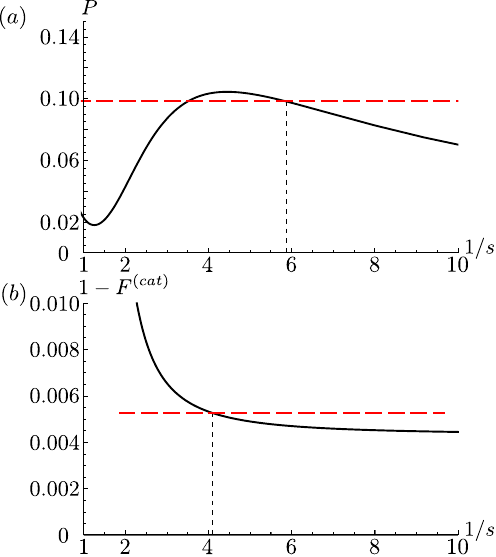}
\caption{Fitting the inverse squeezing parameter $1/s$ of the ancilla oscillator for the cubic phase state-based gate to match the probability density $P=0.098$ [(a), red dashed line], and infidelity $1-F^{\text{(cat)}}=0.005$ [(b), red dashed line] for the Fock state-based gate used here for comparison. The gate using the cubic phase state achieves the same probability density (a), or the fidelity (b) for $\gamma=0.075$, $y_m=2.486$, $s=0.171$, or $\gamma=0.334$, $y_m=11.012$, $s=0.241$, respectively.}
\label{fig_probability_infidelity}
\end{figure}

The gate that uses the cubic phase state provides the same or better probability density for $\gamma = 0.075$, $y_m=2.486$, see Table~\ref{tab_ym_gamma}, within some range of the ancilla coordinate stretching factor $1/s$, as shown in Fig.~\ref{fig_probability_infidelity}(a). The best fidelity $F^{\text{(cat)}}$, such that $1 - F^{\text{(cat)}} = 0.098$, is achieved by the maximal squeezing within this range, which corresponds to $s=0.171$, or $\sim$15 dB of squeezing~\cite{Baeva2023}.

Thus, in this particular example, for the gate with the Fock resource state, the infidelity is as much as by the order of magnitude less than with the cubic phase state. This difference is easily explained in terms of semiclassical visualization. By choosing for mapping an initial coordinate of the target oscillator on the phase plane, represented by a solid blue arrow in Fig.~\ref{fig_Scheme}(a), one shifts the parabola, as shown on the top row of column (c). This changes the spacing between the two cross-sections that represent the measurement procedure and eventually leads to the distortion of copies. As one can observe from~\eqref{cubic_cat_transf} and from the visual representation in Fig.~\ref{fig_Scheme}(c), the distortion of copies due to the parabolic shape of the ancilla support region becomes smaller for larger cubic nonlinearity $\gamma$, so that two branches of the parabola become closer to vertical. The chosen value of the cubic nonlinearity $\gamma = 0.075$ is not enough to meet this criterion.

To achieve the same fidelity as with the Fock resource state, $1 - F^{\text{(cat)}} = 0.005$, one can choose from Table~\ref{tab_ym_gamma} the gate parameters, that correspond to the larger cubic nonlinearity. The first suitable choice is $\gamma=0.334$, $y_{m}=11.012$. Note that a non-perfect squeezing introduces some noise and also may affect the fidelity. For the chosen values of $\{\gamma,\,y_m\}$, the infidelity in dependence on the squeezing parameter is shown in Fig.~\ref{fig_probability_infidelity}(b). The minimal squeezing compatible with the given above value of the fidelity corresponds to $s = 0.241$, or $\sim$12~dB of squeezing. The gain in fidelity is achieved at the expense of the probability density, $P = 0.022$, which is smaller than the corresponding one $P = 0.098$ for the Fock resource state.

\begin{figure*}[t!]
\centering
\includegraphics[width=0.9\textwidth]{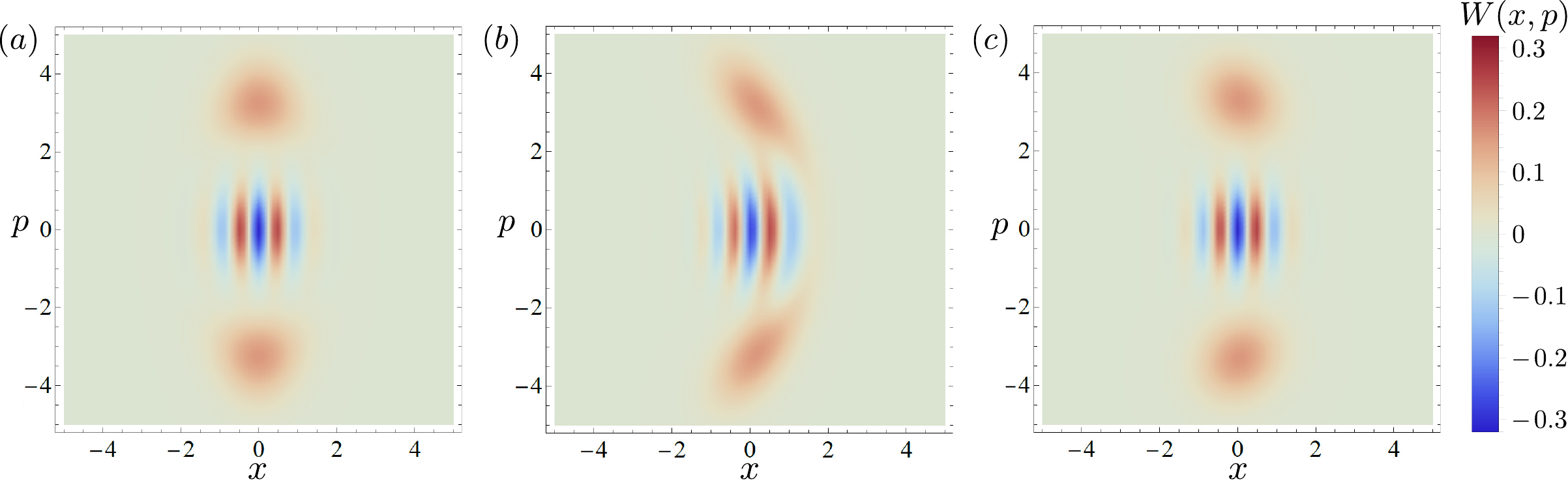}
\caption{Wigner functions of the output cat-like states of the gates that use vacuum input state and the considered above non-Gaussian resource states: (a) the Fock resource state with $n=5$ and optimal measurement outcome $y_m=0$ that corresponds to the odd cat state with the displacement of the copies $\pm p^{(+)}=\pm 3.32$, the probability density $P=0.098$, and the infidelity $1-F^{(cat)}=0.005$; (b) the cubic phase state with the parameters $\gamma = 0.075$, $y_m=2.486$ chosen in such way that the gate has the same probability density of the measurement outcome as in the case (a); (c) the cubic phase state with the parameters $\gamma = 0.075$, $y_m=2.486$ that provide the same infidelity as in the case (a).}
\label{fig_Wigner_comparison}
\end{figure*}
 
The Wigner functions of the output states of the gates that we have compared above are shown in Fig.~\ref{fig_Wigner_comparison}. These plots are in agreement with the qualitative picture and its visualization presented here.

\section{Conclusion}

In this paper, we have shown that the gate using the Fock resource state demonstrates better overall efficiency in producing the cat-like states in terms of fidelity and probability of success in comparison with the gate based on the cubic phase state. The latter one provides a comparable fidelity at the expense of probability, and to be competitive needs rather large cubic nonlinearity, which is evident from a simple geometrical picture. On the other hand, in order to produce the odd/even cat states, the Fock resource states with the odd/even numbers of quanta should be used, which corresponds to different spacing between the copies and could be not suitable for some schemes of error correction. In the case of the resource cubic phase state, a fixed cubic nonlinearity can be used, while the relative phase between the components of the cat state is specified by accepting the suitable measurement outcome of the ancilla momentum~\cite{Masalaeva2022,Baeva2023}. For the gate parameters that provide a comparable fidelity, the spacing between the copies is less dependent on the measurement outcome.

Using our approach to the analysis of different non-Gaussian measurement-assisted gates, one can in some cases easily assess such important features as the distortions introduced by the gate to the output states, the fidelity, the probability of success, etc., with a simple geometrical description based on the semiclassical mapping, including the ``shape'' of the resource state, the entanglement, and the measurement.

\begin{acknowledgments}

A.\:V.\:B. acknowledges a financial support from the Foundation for the Advancement of Theoretical Physics and Mathematics “BASIS” (Grant No. 23-1-5-116-1).

\end{acknowledgments}

\bibliography{Bibliography.bib}
 
\end{document}